\documentclass[conference, 10pt]{IEEEtran}
\usepackage{amsthm,amsmath,amssymb,esint,color,graphicx,overpic,bbm}
\usepackage{tcolorbox}
\newcommand{\subparagraph}{}\usepackage[compact]{titlesec}
\usepackage{cite,hyperref}

\usepackage[ruled,commentsnumbered]{algorithm2e}

\usepackage{algpseudocode}
\usepackage{enumitem}

\theoremstyle{plain}

\theoremstyle{definition}
\theoremstyle{plain}

\theoremstyle{definition}

\providecommand{\definitionname}{Definition}

\providecommand{\lemmaname}{Lemma}
\providecommand{\theoremname}{Theorem}
\providecommand{\remarkname}{Remark}
\newtheorem{definition}{Definition}
\newtheorem{lemma}{Lemma}
\newtheorem{theorem}{Theorem}

\def\QED{\mbox{\rule[0pt]{1.3ex}{1.3ex}}}

\newcommand{\argmax}{\arg\max}

\def\ind{\mathbbm{1}}

\newcommand{\Lc}{{\cal L}}

\newcommand{\Ic}{{\mathcal{I}}}
\newcommand{\Icp}{{\mathcal{I}_+}}
\newcommand{\Yc}{{\cal Y}}
\newcommand{\Xc}{{\cal X}}

\newcommand{\Nc}{{\cal N}}

\newcommand{\norm}[1]{\left\lVert#1\right\rVert}
\IEEEoverridecommandlockouts
\begin{document}
\title{Distributed Popularity Learning for D2D Caching}
\title{Online D2D Caching Policies}
\title{Distributed Online Policies for D2D Caching Networks}
\title{Distributed Online Policies for D2D Caching}
\title{Learning to Cooperate in D2D Caching Networks}

\vspace{-3mm}
\author{\IEEEauthorblockN{Georgios S. Paschos, Apostolos Destounis, George Iosifidis}\\
	\vspace{-4.5mm}
	\IEEEauthorblockA{
	\thanks{{G. S. Paschos, A. Destounis are with the Mathematical and Algorithmic Laboratory, Huawei, France, (georgios.paschos@huawei.com; apostolos.destounis@huawei.com); G. Iosifidis is with Trinity College Dublin.} This work was supported by Science Foundation Ireland, grant 17/CDA/4760.}%
	\vspace{-0.5em}
}}


\maketitle

\addtolength{\floatsep}{-\baselineskip}
\addtolength{\dblfloatsep}{-\baselineskip}
\addtolength{\intextsep}{-\baselineskip}
\addtolength{\textfloatsep}{-\baselineskip}
\addtolength{\dbltextfloatsep}{-\baselineskip}
\addtolength{\abovedisplayskip}{0ex}
\addtolength{\belowdisplayskip}{0ex}
\addtolength{\abovedisplayshortskip}{0ex}
\addtolength{\belowdisplayshortskip}{0ex}
\setlength{\abovecaptionskip}{0ex}
\setlength{\belowcaptionskip}{0ex}

\begin{abstract}
We consider a wireless device-to-device (D2D) cooperative network where memory-endowed nodes store and exchange content. Each node generates random file requests following an unknown and possibly arbitrary spatio-temporal process, and a base station (BS) delivers any file that is not found at its neighbors' cache, at the expense of higher cost. We design an online learning algorithm which minimizes the aggregate delivery cost by assisting each node to decide which files to cache and which files to fetch from the BS and other devices. Our policy relies on the online gradient descent algorithm, is amenable to distributed execution, and achieves asymptotically optimal performance for any request pattern, without prior information.
\end{abstract}

\section{Introduction}

The rapidly growing demand for mobile content delivery \cite{Cisco2015} creates new revenue opportunities for wireless networks, but also requires to increase rapidly their capacity. Unfortunately, typical solutions based on PHY-layer advances or network densification are constantly outpaced by the increasing demand \cite{paschos-jsac}, and this calls for new content delivery approaches. To this end, a potentially game-changing idea is to employ device-to-device (D2D) communications where memory-endowed user devices can cache popular files and exchange them with each other upon request \cite{femtocaching_d2d}. Such cooperative D2D swarms can increase the network content delivery capacity, mitigate  cellular congestion, and improve the end-user experience.  

A key challenge in D2D cooperative caching  is to design the caching policy, i.e., identify which files to cache at each device at any given time \cite{paschos_16}. On the one hand the devices have small storage and therefore can store only a small subset of the possible content files; on the other hand each of them ``sees'' a small number of requests per unit time and hence estimating the popular files at each location becomes very challenging. In view of these limitations, the devices may consistently fail to store in their cache the files that will be requested in the future by their neighbors, rendering the D2D \emph{cache-hit ratio} practically negligible and hence this solution ineffective.

Caching policies often assume that requests are generated by a given stationary process, cf. \cite{paschos-jsac}, and systems in practice rely on \emph{reactive policies} such as LFU and LRU. These, however, solve the 1-cache problem conditionally on the request process. For example, LFU is suitable for stationary requests~\cite{Fricker12}, and LRU for the adversarial model \cite{Mattson70}. When the actual process is other than assumed these policies perform poorly \cite{paschos-infocom19}, and this problem is exacerbated in D2D networks where popularity has hot spots in time and space. Recently, \cite{giovanidis-mLRU, leonardi-implicit, avrachenkov-acm17} proposed dynamic policies for caching networks, e.g., the m-LRU \cite{giovanidis-mLRU} or ``lazy rule'' \cite{leonardi-implicit} policies, which however do not offer performance guarantees. Hence, the problem of designing a policy robust to the D2D network dynamics is an equally challenging and important open problem.

\begin{figure}
	\centering
	\includegraphics[width=0.235\textwidth]{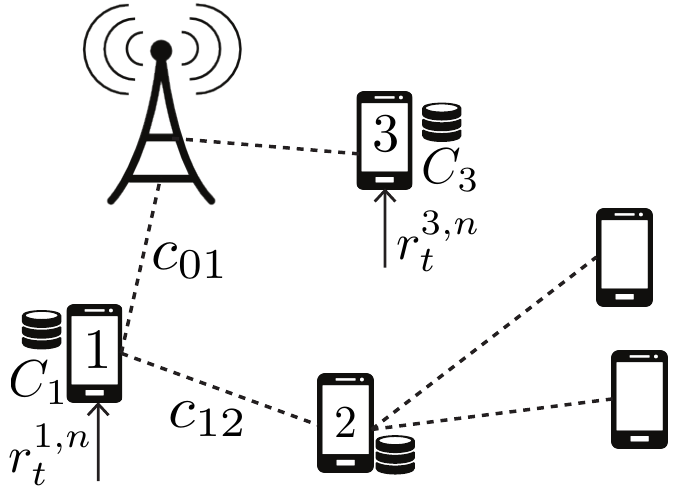}
	\caption{A BS-assisted D2D network. Nodes 1 and 2 can exchange ($\ell_{12}=1$) files with cost $c_{12}$ per byte, but node 3 is only in range with the BS ($\ell_{13}=\ell_{23}=0$). At each slot $t$ a node $i$ can generate a request $r_{t}^{i,n}$ for file $n\!\in\!\Nc$.}
	\label{fig:system}
\end{figure}

In this cooperative D2D network, users might change positions and content preferences, and hence we need an algorithm that will allow each device to decide: \emph{(i)} which files to cache for serving its neighbors' needs; \emph{(ii)} from which neighbors to retrieve a file, and when to use the last-resort solution of the base station (BS), Fig. \ref{fig:system}. This requires a learning mechanism for making caching and routing decisions in an online and decentralized fashion. Previous efforts employing learning are restricted either to inference of the popularity model \cite{Bastug2015TransferExtended, poor-learning-caching}; or rely in Q-learning \cite{giannakis-q-learning, gunduz-reinforcement} and classification techniques \cite{mihaela-video-caching} to estimate request frequencies. These approaches are not applicable to this D2D scenario as they are centralized, exhibit often high complexity, presume a stationary model, and do not decide routing. Other interesting suggestions for D2D networks, see \cite{cao-rl, chen-d2d-learning}, suffer from computational complexity or rely on assumptions that are valid only in some cases, e.g., known popularity. Finally, \cite{C_ioannidis_2010} proposes a decentralized D2D file sharing algorithm which however needs access to the pattern of file requests and network evolution. 

Here we design a caching policy that has \emph{universally-optimal performance}, which is defined as the cost for delivering the requested files to users through (cheap) D2D or (costly) BS-to-device transmissions. We formulate the D2D caching operation as an online convex optimization (OCO) problem, and develop a dynamic and distributed algorithm that solves it without the need to make any assumption about the request pattern. That is, our policy ensures asymptotically \emph{no regret}, as it achieves no more average cost than a static caching configuration selected with knowledge of future requests.

The contributions of this paper can be summarized as follows. \emph{(i)} We propose the idea of embedding a distributed online learning mechanism to D2D caching policies. We achieve this by formulating an OCO problem, and this opens a link between caching and this novel machine learning tool. \emph{(ii)} We design an online caching policy that leverages the online gradient descent algorithm to achieve asymptotically optimal performance under any possible spatio-temporal request pattern in our D2D network. We also explain how our policy can adapt to network changes and user churn. \emph{(iii)} We compare our policy with the state-of-the-art mLRU and ``lazy'' LRU policies, verifying that it outperforms its competitors, while converging to the optimal static policy.


\vspace{1mm}
\section{System model}\label{sec:model}

\textbf{Network}. Consider a set $\Ic$ of $I$ wireless users in an area, each one with a cache of size $C_i\geq 0, i\in \Ic$. Let $\Lc$ be the set of direct links connecting the users, where a link appears if the devices' proximity and the electromagnetic environment allows it to be reliably established. A link between $i$ and $j$ is associated to a cost $c_{ij},\,(i,j)\!\in\! \Lc$, which represents, e.g. application-layer latency performance or energy consumption during content transmission; and we assume $c_{ii}=0, \forall i\in \Ic$. All users maintain a connection with a base station (BS, subscripted with $0$), and we define $\Icp=\Ic\cup\{0\}$ and $I_+=|\Icp|$. The users can obtain any file from the BS at cost $c_{i0}>c_{ij}, ~\forall i\!\in\!\Ic, (i,j)\!\in\!\Lc$. We assume that links can deliver the requested content in the considered time window.

\textbf{Requests}. There is a catalog $\mathcal{N}$ with $N$ files of unit size. The system operation is time-slotted, and $r_{t}^{i,n}\!\in\!\{0,1\}$ denotes the event that a request for file $n$ has been submitted by user $i$ during slot $t$. At each $t$ we assume there is one request, or, from a different perspective, that the system decisions are updated after each request.\footnote{We can also consider batches of requests. If the batch has 1 request from each location, the pattern is biased to equal request rate at each location. An unbiased batch should contain an arbitrary number of requests from each location. Our guarantees hold for unbiased batches of arbitrary finite length.} Hence, the request process is described by a sequence of vectors $\{r_t\}_{t=1,\dots,T}$ drawn from set:
\vspace{-1mm}
\begin{equation}
	\mathcal{R}=\left\{r\in \{0,1\}^{N\!\cdot\! I} ~\bigg |~ \sum_{n\in\mathcal{N}}\sum_{i\in\Ic} r^{i,n}=1\right\}.
\end{equation}
The instantaneous file popularity is expressed by the probability distribution $P(r_t)$ (with support $\mathcal{R}$), which is allowed to be unknown and arbitrary. The same holds for the joint distribution $P(r_1,\dots,r_T)$ that describes the file popularity evolution, for any user location, and within an interval of $T$ slots. This generic model captures all possible spatio-temporal request sequences, including stationary (i.i.d. or otherwise), non-stationary, and adversarial models. The latter is the most general case, as they include request sequences selected by an \emph{adversary} aiming to disrupt the system performance. 

\textbf{Caching}. The cache of each user $i\in\Ic$ can store only $C_i\!<\!N$ files, but the BS has the entire catalog. Following the standard practice in wireless caching models \cite{femtocaching_d2d, paschos-jsac}, we perform caching using the \emph{Maximum Distance Separable} (MDS) codes. In MDS, the files are split into a fixed number of $F$ \emph{data chunks}, and we store in each cache an amount of \emph{coded chunks} that are pseudo-random linear combinations of the data chunks. Using the MDS properties, a user can decode the file (with high probability) if it receives any $F$ coded chunks. Hence, the caching decision vector $y_t$ has $N\!\cdot\!I$ elements, where $y_t^{i,n}\!\in\! [0,1]$ denotes the amount of random coded chunks of file $n$ stored at user $i$ during slot $t$.\footnote{The fractional caching is supported by the observation that large files are composed of thousands chunks, stored independently, see literature of \emph{partial caching}~\cite{maggi}. Hence, by rounding these fine-grained fractional decisions, we will only induce a small application-specific error. In some prior caching models, fractional variables  represent probabilities of caching 
\cite{Shalev12,Blaszczyszyn2014Geographic}.} Based on this, we introduce the convex set of eligible caching vectors: 
\[
\mathcal{Y}=\left\{y\in [0,1]^{N\cdot I} ~\Bigg|~ \sum_{n\in\mathcal{N}}y^{i,n}\leq C_i, ~i\in \Ic\right\}.
\]
We are interested in distributed policies, where each user $i$ changes its cache $y^i_t=( y_{t}^{n,i}, n\!\in\!\Nc)$ based on information from its one-hop neighbors $j:\,(i,j)\!\in\!\Lc$. Thus, we define:
\begin{definition}[Local Caching Policy]
	A local caching policy $\sigma^i$ for user $i$ is a (possibly randomized) rule 
	\[
	\sigma^i \big(r_1,\ldots,r_{t-1}; y_{1}^{i},\ldots,y_{t-1}^{i}\big) : \longrightarrow y_{t}^{i}\in[0,1]^N.
	\]
\end{definition} 
\noindent The collection $\sigma = \{\sigma^i, i\in\Ic\}$  of the caching policies for all users will be henceforth referred to as a ``caching policy". 

\textbf{Routing}. Since each user $i\!\in\!\mathcal{I}$ might have more than one neighbors, we introduce \emph{routing variables} to determine the cache from which  the requested file will be fetched. Let $z^{n,i,j}_t\in[0,1]$ denote the portion of request $r_{t}^{n,i}$ that is fetched from cache $j$, and we define the routing vector $z_t=(z^{n,i,j}_t, n\!\in\!\Nc, i,j\!\in\!\Ic )$ implemented in slot $t$. There are two important remarks here. First, due to the coded caching model, the requests can be simultaneously routed from multiple caches. In terms of communications, this can be implemented through time-sharing among the activated links, or using concurrently different network interfaces. Second, the caching and routing decisions are coupled and constrained: \emph{(i)} a request cannot be routed from an unreachable cache, \emph{(ii)} we cannot route from a cache more data chunks than it has, and \emph{(iii)} each request must be fully routed.

We define $\ell_{ij}\!=\!1$ if $\{(i,j) \in\Lc\}$ and $\ell_{ij}\!=\!0$ otherwise, and thus the set of eligible routing decisions conditioned on $y_t$ is: 
\[
{\cal Z}_t(y_t)\!=\!\left\{z\!\in\![0,1]^{N\!\cdot\! I^2} \Bigg|
\begin{array}{c}
\sum_{j\in \Icp}z^{n,i,j}_t\!=\!r^{n,i}_t, \vspace{1.5mm}\\

z^{n,i,j}_t\!\leq\!\ell_{ij}y^{n,j}, ~i,j\!\in\!\Ic, n\!\in\!\Nc
\end{array}
\right\}.
\]
Note that $z^{n,i,0}_t$ does not appear in the second constraint, because the BS stores the entire catalog and can serve all users. This \emph{last-resort} routing option ensures that ${\cal Z}_t(y_t)$ is non-empty for any $y_t\!\in\! \mathcal{Y}$. As it will become clear next, the optimal routing decisions can be devised for a given cache configuration. This is an inherent property of link-uncapacitated caching networks, see also \cite{femtocaching_d2d, paschos-jsac}.

\vspace{2mm}
\section{Problem Formulation}\label{sec:problem}


A file request of a node can be served, exclusively or partially (due to MDS), by neighboring devices at a smaller cost than fetching it from the base station. Given a cache configuration $y_t$, the (minimum) cost to satisfy $r_t$ is:
\begin{equation}\label{eq:utility}
	f_t(y_t) = \min_{z_t\in{\cal Z}_t(y_t)}\sum_{n\in\Nc}\sum_{i\in\Ic}\sum_{j\in\Icp}r_t^{i, n}c_{ij}z_t^{n, i, j},
\end{equation} 
where the optimization decides the routing that minimizes the cost for a given file placement at the nodes. The function's form suggests that is beneficial if the file has been cached at the device asking for it ($c_{ii}=0$) or at nearby devices that can send it with low cost. However, it is daunting to assess the impact of $y_t$, as it involves the solution of an optimization problem. Fortunately, the cost function above is convex: 
\begin{lemma}\label{lem:convexity}
Function $f_t(y)$ is convex in its domain $\Yc$, $\forall r_t\in{\cal R}$. 
\end{lemma}
\noindent\emph{Proof:} Fix a request vector $r_t$ and consider cache configurations $y_1, y_2\!\in\! \Yc$; note that, for any $\lambda\!\in\![0,1]$, $\overline{y}\!=\!\lambda y_1 \!+\! (1\!-\!\lambda) y_2$ is also a valid configuration. We will show that:
\[
f_t\big(\lambda y_1 + (1-\lambda) y_2\big) \leq \lambda f(y_1) + (1-\lambda)f(y_2)
.\]
Let us denote $z^*_1, z^*_2, \overline{z}^*$ the optimal routing vectors corresponding to $y_1, y_2, \overline{y}$, respectively. We then have: 
\begin{equation}
f_t(y_k) = \sum_{n\in\Nc}\sum_{i\in\Ic}\sum_{j \in \Icp}r^{i, n}_tc_{ij}z^{*, n, i, j}_k, \forall k\in\{1,2\}, \text{and}
\end{equation} 
\begin{align} \nonumber
f_t(\overline{y}) & = \sum_{n\in\Nc}\sum_{i\in \Ic}\sum_{j \in \Icp}r^{i, n}_tc_{ij}\overline{z}^{*, n, i, j}\\ \label{eq:proofConv}
& \leq \sum_{n\in\Nc}\sum_{i\in\Ic}\sum_{j \in\Icp}r^{i, n}_tc_{ij}{z}^{n, i, j}, \forall z\in {\cal Z}_t(\overline{y}).   
\end{align}      
It holds $\lambda z_1^* \!+\! (1\!-\!\lambda) z_2^* \!\in\! {\cal Z}_t(\overline{y})$, thus $\lambda f_t(y_1)\! +\! (1\!-\!\lambda)f_t(y_2)\!=$
\begin{align} \nonumber 
& =\sum_{n\in \Nc}\sum_{i\in\Ic}\sum_{j \in \Icp}r^{i, n}_tc_{ij}\left(\lambda z^{*, n, i, j}_1 + (1-\lambda)z^{*, n, i, j}_2\right) \\  \nonumber 
& \overset{\text{\eqref{eq:proofConv}}}{\geq} f_t\big(\lambda y_1 + (1-\lambda) y_2\big). \,\,\,
\quad\quad\quad\quad\quad\quad\quad\quad\quad\quad\quad\quad\quad\blacksquare
\end{align}

Subscript $t$ at the cost function \eqref{eq:utility} reminds us its dependence on the request $r_t$ that is generated at $t$. Since these events may vary according to a non-stationary process, we will use the concept of regret from online convex optimization \cite{Shalev12}.

We capture that the request sequence may follow any arbitrary and \emph{a priori} unknown probability distribution, by using the idea of an \emph{adversary} which selects $r_t$ at each slot $t$, while knowing $y_t$. This assumption reflects that, in practice, caches are populated before the requests are issued. Since by Lemma \ref{lem:convexity} $f_t(y)$ are convex, our problem falls in the Online Convex Optimization framework \cite{Shalev12}. The performance metric of an algorithm in this line of work is the \emph{regret}: the difference between costs incurred by the algorithm and the \emph{best static configuration in hindsight}. In our case, this benchmark is the optimal cache configuration (same for all slots) devised with knowledge of all requests in the time horizon of interest $T$. Hence, the regret of policy $\sigma$ is: 
\begin{equation}\label{eq:regret}
R_T(\sigma) = \max_{P(r_1,r_2,..r_T)}\mathbb{E}\left[\sum_{t=1}^T f_t(y_t(\sigma)) - \sum_{t=1}^Tf_t(y^*)\right]
.\end{equation} 
The expectation is over the joint probability distribution of requests and possible randomizations in $\sigma$ and,
\[
y^* \in \argmax_{y\in\Yc}\sum_{t=1}^Tf_t(y),
\] 
is the best fixed action in hindsight, i.e. the best chunk placement over the entire sample path of requests. Our goal is to devise a policy whose regret scales sublinearly with $T$:
\begin{equation}
\lim_{T\rightarrow \infty}{R_T(\sigma)}/{T}= 0. \nonumber
\end{equation} 

This ``no regret" property implies that the algorithm learns to perform as good as the best cache configuration $y^*$. Note that if the requests are i.i.d. ``no regret" implies that the performance of the policy approaches the optimal in terms of $\sim_tf_t(y^*)$. However, our adversarial model is much more general; in this case, comparing to a static policy is a way to limit the power of the adversary while still being able to obtain meaningful policies, which are robust for all request models. 

\vspace{2mm}
\section{Distributed D2D Caching Algorithm}\label{sec:algorithm}
		
Our distributed caching algorithm is based on online gradient descent \cite{Zi03}. The main idea is to use the first order approximation $f(y) = f_t(y_t) + \nabla f_t(y_t)^\top(y - y_t)$ as a predictor of the unknown function $f_{t+1}(y)$ that the adversary will select next. The caching configurations, then, are updated by taking an appropriate step in the direction of the gradient $\nabla f_t(y_t)$.

\subsection{Finding the Direction of Improvement}
Since the utility function \eqref{eq:utility} is not necessarily differentiable everywhere, we will rely on subgradients. In order to find one, we first simplify $f_t(y)$. Let us denote $i_t, n_t$ the user making the request and the file requested at slot $t$, respectively, and $\mathcal{J}(i)=\{j\in{\Icp}: \ell_{ij}=1\}$ the set of the nodes (including the BS) connected to user $i$. Then, it is $z^{n,i,j} = 0, \forall n\neq n_t, i\neq i_t, j\notin \mathcal{J}(i_t)$, and hence $f_t(y_t)$ simplifies to:
\begin{align}\label{eq:simplifiedUtility}
f_t(y_t) = & \min_{z\in [0,1]^{I_{+}}}\sum_{j\in\mathcal{J}(i_t)}c_{i_tj}z^{n_t, i_t,j} \\
\text{s.t. } & \sum_{j\in\mathcal{J}(i_t)}\!\!\!z^{n_t, i_t,j} = 1 \label{eq:cstr} \\ 
			 & z^{n_t, i_t,j} \leq y^{n_t, j}_t,\,\,\,\,\,\, \forall j \in \mathcal{J}(i_t)\,.\label{eq:cstr2}
\end{align}
Equations \eqref{eq:simplifiedUtility}-\eqref{eq:cstr2} define an optimization problem, henceforth referred to as $P_f$, the solution of which yields the optimal routing for any $y_t$ (constant input for $P_f$). That is, to evaluate $f_t(\cdot)$ at vector $y_t$ we need to solve $P_f$. Despite this intricate form of $f_t(y_t)$, we show that it is possible to obtain a subgradient which is needed for our online caching algorithm. 

We first define the Lagrangian of $P_f$ as follows:
\begin{align} \nonumber 
L_t(y, z, \alpha, \beta) &= \sum_{j\in\mathcal{J}(i_t)}c_{i_tj}z^j + \alpha\left(\sum_{j\in\mathcal{J}(i_t)}z^j - 1\right)\\
& + \sum_{j\in\mathcal{J}(i_t)}\beta^j(z^j - y^{j}),
\end{align}
where $\alpha$ and $\beta^j$ are the dual variables, and we simplified notation by dropping $n_t, i_t$. We will prove that the subgradient of $f_t$ at $y_t$ is the optimal dual variables for \eqref{eq:cstr2} in $P_f$.  

\begin{lemma}[Subgradient]
Let:
\begin{equation}\label{eq:subgr}
\beta^*(y_t) \in \argmax_{\beta \geq 0}\left(\max_{\alpha\in\mathbb{R}}\min_{z\in[0,1]^{I_+}}L(y_t, z, \alpha, \beta)\right)
\end{equation}
and define:
\[
g^{n,i,j}_t(y_t) = \begin{cases}
-\beta^{*,j}(y_t), \text{ if } i=i_t, n=n_t, j\in\mathcal{J}(i_t)\\
0, \text{ otherwise}
\end{cases}
.\]
	Then $g_t(y_t)$ is a subgradient of $f_t$ at $y_t$, that is: $f_t(y')\geq f_t(y_t) + g_t(y_t)^\top(y' - y_t), \forall y' \in \Yc$. 
\end{lemma} 
\emph{Proof}. We start by denoting  $\beta^*(y')$ the outcome of \eqref{eq:subgr} for cache configuration $y'\!\in\!\Yc$ and define the function: 
\vspace{-1mm}
\begin{equation}
\Lambda(y, \beta) = \max_{\alpha\in\mathbb{R}}\min_{z\in[0,1]^{I_+}}L(y, z, \alpha, \beta)	
, \forall y \in \Yc,\,\text{and hence:}\nonumber
\end{equation}
\begin{align} \nonumber 
	f_t(y_t) & \overset{(a)}{=} \Lambda(y_t, \beta^*(y_t)) \overset{(b)}{=} \Lambda(y', \beta^*(y_t)) + \beta^*(y_t)(y'-y_t) \\ & \nonumber \overset{(c)}{\leq} \Lambda(y', \beta^*(y')) + \beta^*(y_t)(y'-y_t)
,\end{align}
where ($\alpha$) holds since \eqref{eq:simplifiedUtility}-\eqref{eq:cstr2} has the strong duality property; (b) holds since $\Lambda(\cdot)$ is linear and we can maximize successively over the different primal or dual variables;  and (c) holds as only $-\sum_{j\in\mathcal{J}(u_t)}\beta^jy^{n_t, j}$ in $L_t(\cdot)$ depends on $y$. Due to strong duality for $y'$, we can replace $f_t(y')\!=\!\Lambda(y', \beta^*_t(y'))$, and then suffices to rearrange terms. $\blacksquare$

Since $\beta^*_t(y_t)$ is the optimal multiplier for \eqref{eq:cstr2}, it has nonzero elements only where this constraint is tight. Intuitively, this means that after user $i_t$ requests file $n_t$, the direction of the subgradient is towards caching more parts of this file at user $i_t$ and at users having low-cost D2D links with $i_t$.  

\subsection{Algorithm Design} 

The Distributed Online Caching Policy (DOCP) is shown in Algorithm 1. The execution of the policy is iterative, where in each slot $t$ the following steps take place. First, a user $i_t$ submits a request for a file $n_t$ (step 3). This user solves \eqref{eq:simplifiedUtility}-\eqref{eq:cstr} to find the optimal routing for the current caching configuration $y_t$ (step 4), and requests the parts of $n_t$ from the respective neighbors or the BS (step 5). A certain utility is accrued based on this routing $z_{t}^*$ and the existing $y_{t}^*$ (that was calculated based on previous requests). Then, user $i_t$ sends the optimal multiplier $\beta^{*,j}_t(y_t)$ to each neighbor $j\in\mathcal{J}(i_t)$ (step 8) who updates its caching policy accordingly. This involves calculating the new $y_{t+1}^*$, based on the latest request, and projecting them back into the feasible space (step 9):
\begin{equation}
y_{t+1}^{j} = \Pi_{[0,1]^N}\left(y_t^i + \gamma \beta^{j, *}_t(y_t)e_{n_t}\right), \nonumber
\end{equation}
where $\Pi_{\Xc}(.)$ is the Euclidean projection on $\Xc$, and $e_n\!\in\!\mathbf{R}^N$ has zero elements except the $n$-th element being equal to 1.

Note that DOCP is indeed distributed, since only the neighbors of each requester need to update their caches. Moreover, this update is based solely on messages received by the requester, and these communication overheads are moderate as only the Lagrange multipliers are sent to 1-hop neighbors. Finally, the projection operation can be executed efficiently, i.e., in $O(N\log N)$ runtime, and for each user independently, by using the local projection algorithm introduced in \cite{paschos-infocom19}. We omit the details here due to lack of space.

\subsection{Performance Guarantees}
The next theorem proves that DOCP achieves no regret performance, under any possible spatio-temporal arrival pattern.
\vspace{1mm}
\begin{theorem}[Regret of DOCP]
	For step size $\gamma = \frac{\sqrt{2CJ^*}}{c^*\sqrt{T}}$, the regret of DOCP satisfies: 
	\[
	R_T(DOCP) \leq c^* \sqrt{2CJ}\sqrt{T}
	,\]
	where, we defined the parameters $c^*\!=\!\max_{i \in \Ic}c_{i,0}$, $C\!=\!\max_{i\in\Ic}C_i$ and $J^*\!=\!\max_{i\in\Ic}|\mathcal{J}(i)|$. 
\end{theorem}
\emph{Proof:} Using non-expansiveness of Euclidean projection: 
	\begin{align}\nonumber 
	&\norm{y_{t+1}-y^*}^2\leq \norm{y_t - \gamma g_t(y_t) - y^*}^2\\ \nonumber 
	 & =\norm{y_t-y^*}^2 + \gamma^2\norm{g_t(y_t)}^2 - 2\gamma g_t(y_t)^\top (y_t-y^*)   
	.\end{align}
Also, a telescopic sum over $T$ slots gives $\norm{y_{T}-y^*}^2\leq$
	\[ 
	\leq\norm{y_{1}-y^*}^2 + \gamma^2\sum_{t=1}^T\norm{g_t(y_t)}^2 - 2\gamma \sum_{t=1}^Tg_t(y_t)^\top (y_t-y^*) 
	.\]
	To proceed, note that $\norm{y_1 - y^*}^2\leq 2CJ^*$ and $\norm{g_t(y_t)}\leq (\max_{i\in\Ic}c_{0,i})$. Using that $\norm{y^T-y^*}^2\geq 0$ and rearranging: 
	\[
   	\sum_{t=1}^Tg_t(y_t)^\top(y_t - y^*) \leq \frac{{2CJ^*}}{2\gamma} + \frac{\gamma T (c^*)^2}{2} 	
	.\]
	Furthermore, due to convexity of $f_t(y)$, it holds $f_t(y^*) - f_t(y_t) \geq g_t(y_t)^\top(y^* - y_t)$, thus: 
	\[
	R_T(DOCP) = \sum_{t=1}^T(f_t(y_t) - f_t(y^*) ) \leq \frac{{2CJ^*}}{2\gamma} + \frac{\gamma T (c^*)^2}{2}
	.\]
	The value of the step size and regret bound follow by minimizing the Right Hand Side of the above inequality. $\blacksquare$

\begin{algorithm}[t]
	\small
	\nl \textbf{Input:} Select step size $\gamma$\\%
	 \nl \For{$t=1,2,\ldots,T$}{
	 \nl User making request: $i_t$; file requested: $n_t$. \\%
	 \nl User $i_t$ solves \eqref{eq:simplifiedUtility} - \eqref{eq:cstr2} to find $z_t^*=\big(z_t^{*, n_t,i_t,j},j\!\in\!\mathcal{J}(i_t)\big)$.\\%
	 \nl User $i_t$ fetches data from BS or D2D links, based on $z_t^{*}$. \\%
	 \nl Utility $f_t$ is accrued based on $y_{t}^*$, $z_{t}^*$. \\%
	 \nl New caching configuration is devised:\\
	 \For{$j=1,\ldots, |\mathcal{J}(i_t)|$}{
		\nl User $i_t$ sends to user $j$ message $\beta^{*,j}_t$ using \eqref{eq:subgr}. \\%
		\nl User $j$ updates its cache by $y_{t+1}^{j} = \Pi_{[0,1]^N}\left(z^j\right)$,
where $z^{j, n} = y_t^{j, n} + \gamma \beta^{*,j}_t\ind{\{n = n_t\}}, \forall n\in\Nc$.\\%
	} 
	}
\caption{Distributed Online Caching Policy (DOCP)} 	\label{alg:DistrAlg}
\end{algorithm} 
\normalsize

Hence, no regret is achieved with a constant step which depends on $T$, and if $T$ is unknown we can select step $1/\sqrt{t}$, or employ the \emph{doubling trick}, see \cite[Sec. 2.3]{Shalev12}. 

\subsection{Dynamic Network Costs}

The above model and analysis can be readily extended for the case where users change positions in different slots; or the user population evolves with time; or, finally, the users are static but the channel conditions vary. In particular, these scenarios can be captured by the updated cost function:
\begin{equation}\label{eq:utility2}
		f_t(y_t) = \min_{z_t\in{\cal Z}_t(y_t)}\sum_{n\in\Nc}\sum_{i\in\Ic}\sum_{j\in\Icp}r_t^{i, n}c_{ij}^tz_t^{n, i, j},
\end{equation}
where we have replaced the previously constant link costs with slot-specific ones $\{ c_{ij}^t,\,\forall (i,j)\in\mathcal{L}, t=1,\ldots,T \}$. For instance, if link $(i,j)$ exists in slot $t$ but not in slot $(t+1)$ (e.g., nodes have moved farther), then we can use $c_{ij}^{t+1}=c_{max}>\max\{c_{i0},\forall i\in \Ic\}$ which will make this link non-eligible for DOCP (the BS is available and cheaper). It is interesting to note that this extension does not change the regret bound, which is set by the highest cost of the available links, that remains the one between any device and the BS.

\vspace{2mm}
\section{Numerical Results}\label{sec:Simulations}

We illustrate the performance of DOCP in a setting with $N=100$ files, and $8$ devices which are equipped with a cache of capacity $C=6$ and are placed randomly in a cell of size $1.5$Km. Devices can communicate if they are within a range of $500$m, as in current LTE-direct standards \cite{HausSurvey}. The (relative) cost of downloading a file from the base station is set to $10$; a device can fetch a file from its cache at no cost; and the respective costs from other devices vary with the distance: $2$ if the device is closer that $100$m, $5$ if the distance is within $[100m, 300m)$, $7$ if in $[300m, 400m)$, and $9$ if in $[400m, 500m]$. File requests are drawn from a power law distribution with exponent $0.9$. We compare DOCP with the best static policy in hindsight and the lazy LRU and mLRU.

\begin{figure}
	\centering
	\includegraphics[width=0.27\textwidth]{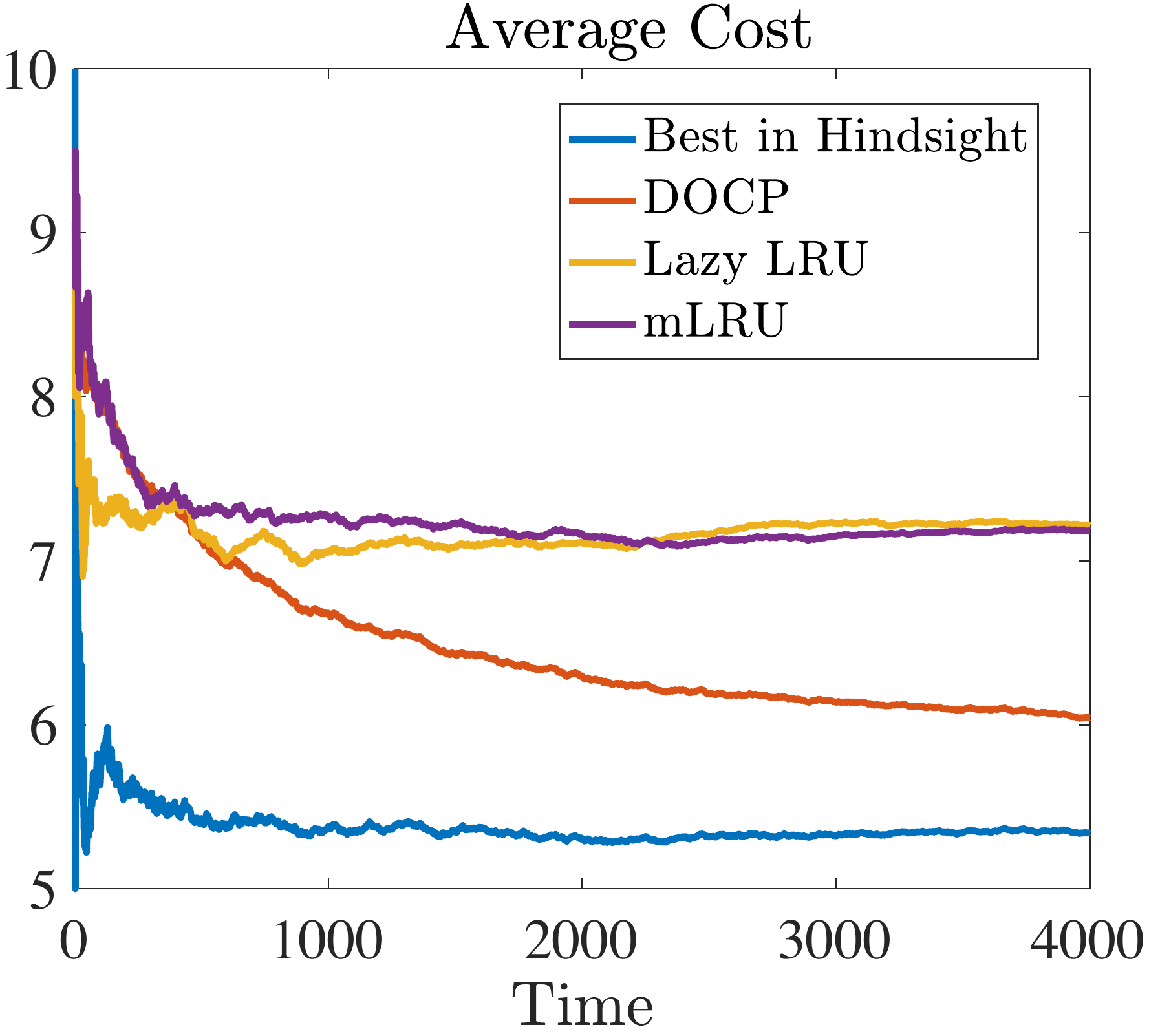} 
	\caption{Comparison of running averages of costs obtained by DOCP, and two competitor reactive policies. The plot presents also the best static policy.}
	\label{fig:simulationExample1}
\end{figure}

\begin{figure}
	\centering
	\includegraphics[width=0.275\textwidth]{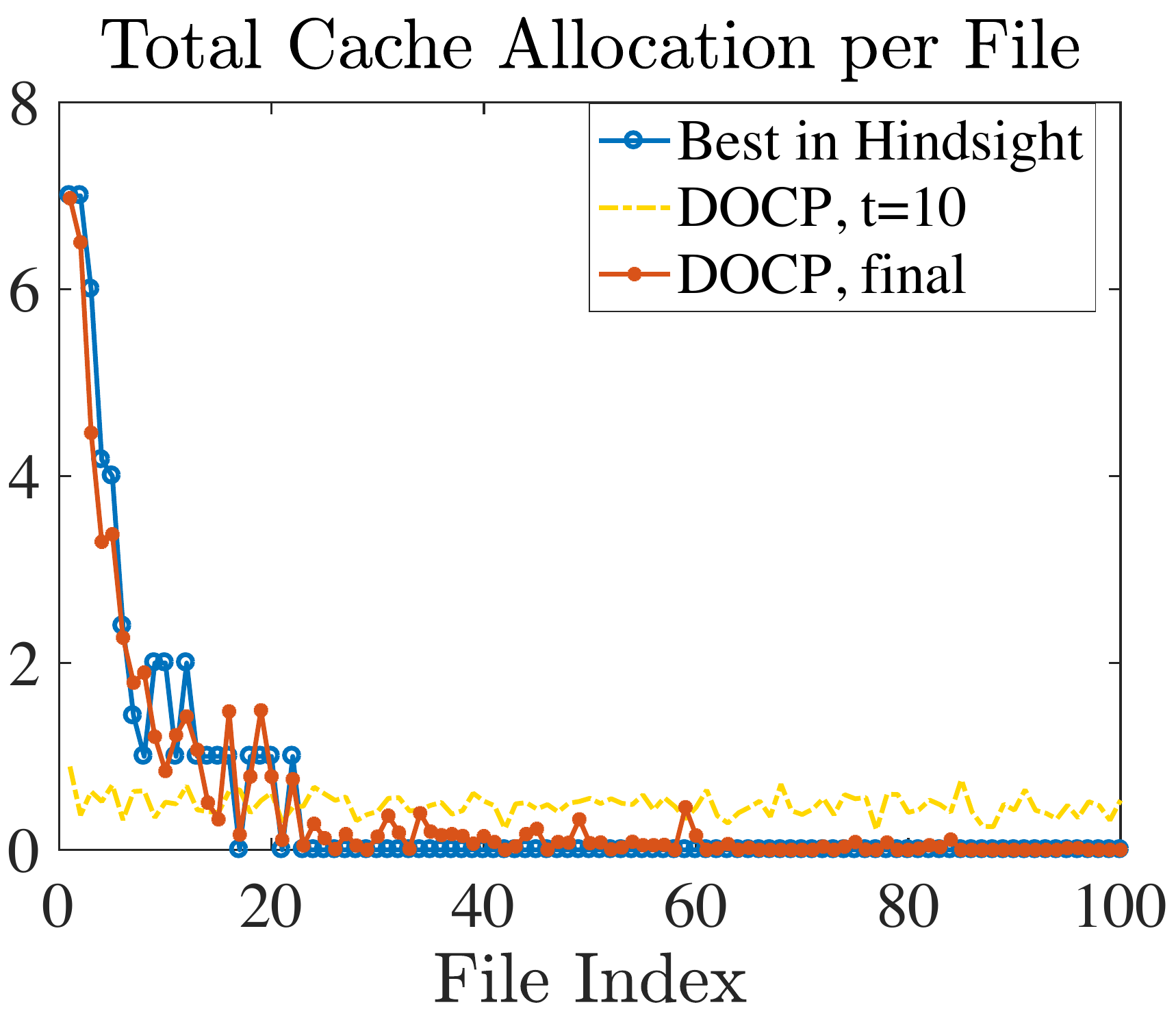}
	\caption{Total cache capacity allocated per file for DOCP, at $t\!=\!10$ and at the last slot, and the best static allocation in hindsight.}
	\label{fig:simulationExample2}
\end{figure}

Our results are presented in Fig. \ref{fig:simulationExample1} which shows the empirical average of the cost for the different policies. We observe that DOCP outperforms both LRU and mLRU, and the margin gets wider as time progresses. In addition, the performance of DOCP gets closer to the one of the best policy in hindsight, thus verifying the no-regret theoretical guarantee. Figure \ref{fig:simulationExample2} compares the total cache allocation, i.e., the total fraction of each file cached at the devices, for DOCP and the best static hindsight policy. We  see that, while the algorithm starts from an almost uniform allocation, by the end of the time interval $T=4000$ the DOCP cache contents are very aligned with the best configuration in hindsight. This demonstrates that DOCP indeed tends to learn the best static configuration. 


\vspace{1mm}
\section{Conclusions}\label{sec:conclusions}

D2D cooperative caching is certainly very promising, but raises previously unseen challenges in devising effective caching policies. Here, we used OCO, a fast-developing area of machine learning, to design an online distributed caching and routing policy that adapts to any (unknown) spatio-temporal request process. This makes it an ideal candidate for such dynamic, often sparse, caching networks. Our work opens a new exciting area at the nexus of online learning and D2D caching systems, and a fascinating next step is to explore how such mechanisms can incorporate incentives for ensuring users' cooperation, leveraging credit mechanisms \cite{iosifidis-upns} or the human tendency to build reciprocal sharing relationships \cite{iosifidis-natcom}.

\vspace{0.5mm}
\bibliographystyle{abbrv}
\bibliography{mybib_v13}
\end{document}